# REDUCED-ORDER MODELLING OF THE BENDING OF AN ARRAY OF TORSIONAL MICROMIRRORS

*A. Molfese[1,2], A. Nannini[2] and F. Pieri[2]*
[1] *IEIIT, C.N.R. - Sezione di Pisa, Italy*
[2] *Dipartimento di Ingegneria dell'Informazione, Università di Pisa, Italy*

**ABSTRACT**

An array of micromirrors for beam steering optical switching has been designed in a thick polysilicon technology. A novel semi-analytical method to calculate the static characteristics of the micromirrors by taking into account the flexural deformation of the structure is presented. The results are compared with 3D coupled-field FEM simulation.

## 1. INTRODUCTION

Several MEMS torsional micromirrors for optical switching have been presented [1-3], most of them fabricated by using bulk micromachining, others with surface micromachining, but requiring delicate assemblage steps after releasing [2]. As it is important to balance device performance and scalability with the maturity, reliability, and manufacturability of the process, a big effort is made in the direction of avoiding these steps and some solutions, like self-assembly [3-4] have been proposed. A robust solution should also avoid multi-wafer assemblies without limiting the angular dynamics of the micromirror. This is not readily obtainable in thin-film surface micromachining technologies because of the small gaps between moving and fixed parts. In the mirror proposed in this work, this problem is overcome by substituting a single micromirror with an array of smaller phased micromirrors, a structure resembling a Venetian blind (Fig. 1).

The characteristics of torsional micromirrors have been studied extensively [5-8]. Like most electromechanical Microsystems, micromirrors show an instable behaviour (the pull-in) above a critical deflection angle, when the electrostatic force/torque overcomes the mechanical force/torque and the movable plate of the micromirror snaps abruptly to the fixed electrode plate. The pull-in parameters, namely, pull-in angle, pull-in voltage, and pull-in displacement, define the maximum performances of the micromirror. The pull-in parameters are determined by the geometrical design of the micromirror and actuating electrodes. In order to design these structures, a accurate model of the expected

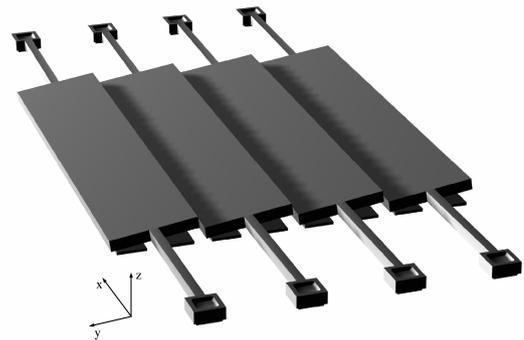

Figure 1: Simplified structure of the array of phased micro-mirrors and underlying actuation.

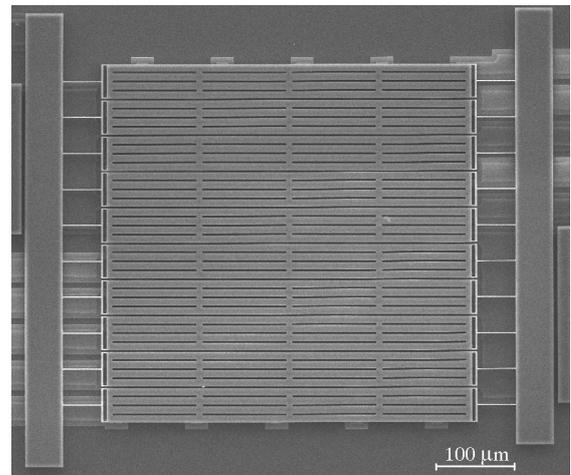

Figure 2: SEM Micrograph of designed micromirrors.

characteristics as a function of the design parameter is necessary.

FEM Full coupled-field analysis of such structures is CPU-intensive and limited by convergence problems. Analytical prediction of the voltage/tilt relationship for such structures is therefore of great aid to the designer. Common simplifying assumptions to make the problem tractable are supposing a rigid vertical translation [5-7] or neglecting vertical deflection altogether[8]. We present here a self-consistent, computationally fast approach to the problem, which also takes into account the vertical deflection and bending of the mirror strip, allowing the calculation of its deformed shape. The obtained results are then compared with FEM simulations.

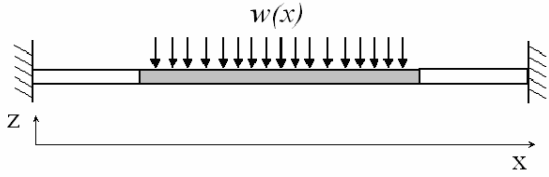

Figure 3: Transversal section of a micromirror.

Finally the applicability of the method to micromirrors of different size is exploited.

## 2. DESCRIPTION OF THE ARRAY

A two-level, thick polysilicon, surface micro-machining process, made available by STMicroelectronics, is used for the mirrors [9]. CMP polishing ensures the micromirror flatness, while the thickness increases its bending stiffness. Long longitudinal holes in the mirror are required for proper release of the moving parts. This can be done without excessive degradation of the optical characteristics. The Venetian blind is constituted by ten micromirrors (Fig. 2), each suspended by two torsional springs. The springs are 50 µm long; the full micromirror array is about 490 µm x 490 µm in size.

Underlying electrodes, fabricated in the first polysilicon layer, are used for actuation. Their width was chosen non-uniform as a compromise between angular dynamics (which is reduced by pull-in for wider electrodes [7]) and technological constraints which required a specific minimum width.

## 3. COMPUTATIONAL APPROACH

If a voltage $V$ is applied between the mirror and one of the underlying electrodes, the distributed electrostatic force yields a torque which rotates the mirror, but also produces a deformation as the structure bends under the load. For ease of description, we suppose here a constant width electrode (Fig. 4), even if the actual computation was performed with the variable width geometry. Let $x$ be the coordinate along the mirror (Fig. 4), $w(x)$ the electrostatic force per unit length and $u_z(x)$ the vertical displacement of the central axis of the mirror. The capacitance $c_m$ per unit length between the mirror and the electrode can be written by using a parallel-plate approximation, and is a function of the tilt angle $\theta$ and of $z$.

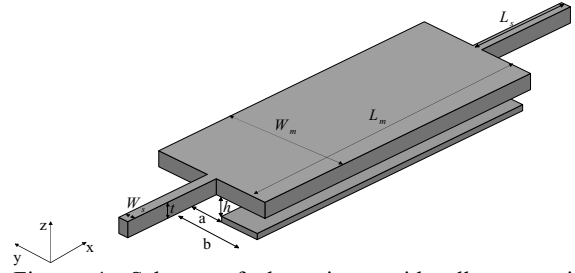

Figure 4: Scheme of the mirror with all geometrical parameters.

Our approach is based on the hypothesis that, for small bending, the distributed load $w(x)$ can be considered equivalent to a uniform load $w_{eq}$. This is clearly much less stringent than supposing a uniform $u_z$ [6-7]. Elementary beam theory gives a closed expression (a polynomial in x) for $u_z(x; w_{eq})$ under a uniform load. Hence, a closed expression for the capacitance $c_m(\theta, z(w_{eq}))$ exists.

To compute the angle-voltage curve the following algorithm is used:

1. Choose rotation angle $\theta$
2. At the first step impose $w_{eq,0}=0$
3. At each step i, calculate $V_i(\theta)$ from the static torsional equilibrium equation:

$$\frac{V_i^2}{2} \int_{mirror} \left.\frac{\partial c_m(\varphi, z(w))}{\partial \varphi}\right|_{\substack{\varphi=\theta \\ w=w_{eq,i}}} dx = k_\theta \cdot \theta \quad (1)$$

4. Compute the equivalent distributed load $w_{eq,i+1}$ from:

$$w_{eq,i+1} \cdot L_m = \frac{V_i^2(\theta_0)}{2} \int_{L_m} \left.\frac{\partial c_m(\varphi, z(w))}{\partial z}\right|_{\substack{\varphi=\theta \\ w=w_{eq,i}}} dx \quad (2)$$

5. Go to step 3 until $|w_{eq,i+1}-w_{eq,i}|$ is less than the chosen tolerance.

In eq. (1) $k_\theta$ is the elastic torsional constant of the springs[10]:

$$k_\theta = 2\frac{GJ_p}{L_s} \quad (3)$$

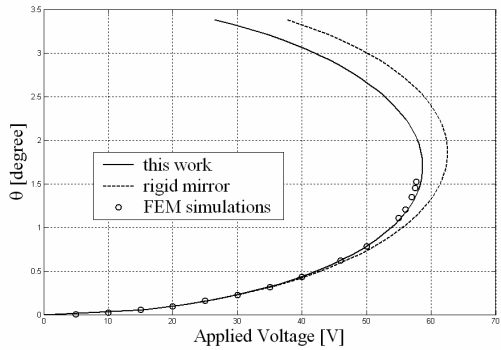

Figure 4: Angular rotation vs. applied voltage of one of the designed micromirrors.

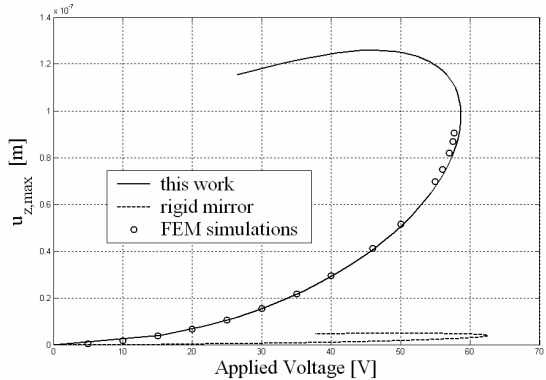

Figure 5: Maximum vertical displacement vs. applied voltage of one of the designed micromirrors.

$G$ is the shear modulus of the polysilicon, $L_s$ is the spring length, $J_p$ is the torsion constant of the rectangular cross-section [10]:

$$J_p = \frac{1}{3} t W_s^3 \left( 1 - 192 \frac{t}{W_s \pi^5} \sum_{i=1,3,5}^{\infty} \left( \frac{1}{i^5} \tanh\left( \frac{\pi W_s i}{2t} \right) \right) \right) \quad (4)$$

The expression of the first derivatives of the capacitance can be analytically derived from the parallel-plate formula, giving:

$$\left. \frac{\partial c_m(\theta, z)}{\partial \theta} \right| = \frac{\varepsilon_0}{\theta^2} \left( \ln\left( \frac{h - b\theta}{h - a\theta} \right) + \frac{h}{h - b\theta} - \frac{h}{h - a\theta} \right) \quad (5)$$

$$\left. \frac{\partial c_m(\theta, z)}{\partial z} \right| = \frac{\varepsilon(b-a)}{(h - b\theta - z)(h - a\theta - z)} \quad (6)$$

The closed expression for $u_z(x; w_{eq})$ under a uniform load is obtained by solving the classic Euler beam equation system with the appropriate boundary conditions[10]:

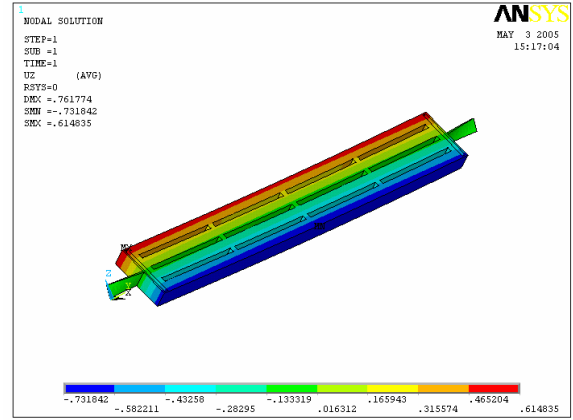

Figure 6: Example of the deformed shape obtained from ANSYS simulations

$$\begin{cases} \dfrac{\partial^4 u_z(x)}{\partial x^4} = 0 & x \in springs \\ \dfrac{\partial^4 u_z(x)}{\partial x^4} = \dfrac{w_{eq}}{EI_2} & x \in mirror \end{cases} \quad (7)$$

where $E$ is the Young's Modulus of polysilicon and $I_2$ is the moment of inertia of the mirror section. An implicit assumption of our mechanical model is that every section rotates rigidly around the $x$ axis. This is certainly reasonable if the mirror width is much smaller than its length.

The algorithm returns, for each $\theta$, the value of the voltage $V(\theta)$ and the equivalent uniform distributed load $w_{eq}(\theta)$. From these values the vertical deformation $u_z(x)$ is deduced.

## 4. RESULTS AND DISCUSSION

The computed $\theta(V)$ is shown in Fig. 4. In Fig. 5, the maximum vertical displacement $u_{z,max}$ (in the central section of the micromirror) is also plotted (bottom). The computation of the full curves requires less than 3 seconds on a PC.

For comparison, the rigid mirror model [5-6] and a full FEM analysis have been performed, and the obtained results are shown in Fig.4 and in Fig. 5 as well. The selected FEM approach (Fig. 6) (the "essolv" ANSYS® macro) required 30 minutes or more for each point in the $\theta(V)$ curve on the same PC.

The proposed approach clearly represents a large improvement with respect to the commonly reduced-order models, with a similar computational burden, but also shows a very good agreement with FEM results. This last result, however, depends on a few assumptions, which are easily verified for the studied geometry, but can fail in more general cases.

Specifically, a first important condition is that the bending of the beam is small enough to ensure that the distributed electrostatic force along the beam can be approximated with the equivalent uniform load. As the electrostatic force is a non-linear function of the distance between the beam and the fixed electrode, it is required the maximum deflection of the beam (i.e. at its center) is much smaller than the gap between it and the fixed electrode. In our case, the maximum deflection is about *0.1 µm* at pull-in (see Fig. 5), while the gap is *1.6 µm*, so that this condition is certainly verified.

A more compelling evidence that the uniform and non-uniform load can be considered equivalent is given in Fig. 7. In the graph, the deformed shape of the central axis of the beam due to the distributed load (as calculated by ANSYS) and the same shape as computed by our method are compared. Three different deflected shapes, for 30, 40, and 50 V of bias (corresponding to rotation of 0.23, 0.44 and 0.78 degrees, respectively) are shown. The error is less than 2% in the worst case.

Another already mentioned assumption is that every section of the mirror is not deformed (i.e. remains rectangular) under the applied load. In our case, the mirror is about 10 times longer than it is wide, and thus has a much larger flexural rigidity along *x* than along *y*, so that this assumption looks reasonable.

## 5. CONCLUSIONS AND FUTURE WORKS

A new method for the fast prediction of the static characteristic of an electrostatically actuated torsional micromirror, taking into account the bending caused by the electrostatic force, was presented. The method was validated against FEM simulations, which showed a very good agreement with the model. An experimental validation of the method by measurement of the pull-in voltage of polysilicon micromirror arrays, as well as a verification of the its applicability to more general geometries are planned.

## 6. ACKOWLEDGEMENTS

The authors would like to thank *STMicroelectronics* for the fabrication of the devices. This work was supported by the Italian MIUR project PRIN "*Matrix of polysilicon micromirrors for optical switching*"

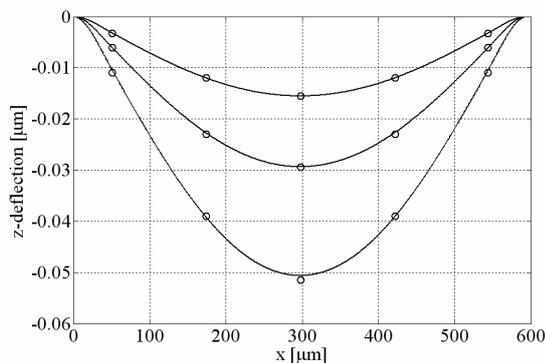

Figure 7: Comparison between the deformed shape of the central axis obtained from the analytical model (solid line) and from ANSYS simulations (circles).